\documentclass[rapids]{jfm}
\usepackage{graphicx}
\usepackage{epstopdf, epsfig}
\usepackage[utf8]{inputenc}
\usepackage{natbib}
\usepackage{mathrsfs,gensymb}
\usepackage{amsmath,amssymb}
\usepackage{setspace,xcolor}
\usepackage{lscape,comment,multirow,appendix}
\usepackage{maths_commands}
\usepackage{color}
\usepackage[normalem]{ulem}

\newcommand{\rev}[1]{{\color{black}#1}}
\newcommand\revout{\bgroup\markoverwith{\textcolor{blue}{\rule[0.5ex]{2pt}{1pt}}}\ULon}

\newcommand{\pg}[1]{{\color{red}#1}}

\newcommand\kssout{\bgroup\markoverwith{\textcolor{green}{\rule[0.5ex]{2pt}{1pt}}}\ULon}
\newcommand\pgout{\bgroup\markoverwith{\textcolor{red}{\rule[0.5ex]{2pt}{1pt}}}\ULon}
\newcommand\gpcout{\bgroup\markoverwith{\textcolor{teal}{\rule[0.5ex]{2pt}{1pt}}}\ULon}
\newcommand\cpcout{\bgroup\markoverwith{\textcolor{magenta}{\rule[0.5ex]{2pt}{1pt}}}\ULon}
\newcommand\lcout{\bgroup\markoverwith{\textcolor{orange}{\rule[0.5ex]{2pt}{1pt}}}\ULon}

\shorttitle{Numerical validation of scaling laws for stratified turbulence}
\shortauthor{Garaud et al.}

\title{Numerical validation of scaling laws for stratified turbulence}

\author{Pascale Garaud\aff{2}, Gregory P. Chini\aff{3}, Laura Cope\aff{4},  Kasturi Shah\aff{1,5} and Colm-cille P. Caulfield\aff{6,1}\corresp{\email{cpc12@cam.ac.uk}}}
\affiliation{
\aff{1} Department of Applied Mathematics and Theoretical Physics, University of Cambridge,
Cambridge CB3 0WA, UK
\aff{2} Department of Applied Mathematics, Baskin School of Engineering, University of California Santa
Cruz, Santa Cruz, CA 95064, USA
\aff{3} Program in Integrated Applied Mathematics and Department of Mechanical Engineering, University of New Hampshire, Durham, NH 03824, USA
\aff{4} School of Mathematics, University of Leeds, Leeds, LS2 9JT, UK
\aff{5} Department of Earth, Atmospheric and Planetary Sciences, Massachusetts Institute of Technology, Cambridge MA 02139, USA
\aff{6} Institute for Energy and Environmental Flows, University of Cambridge, Cambridge CB3 0EZ, UK
}

\begin{document}

\maketitle

\begin{abstract} 
Recent theoretical progress using multiscale asymptotic analysis has revealed various possible regimes of stratified turbulence. Notably, buoyancy transport can either be dominated by advection or diffusion, depending on the effective P\'eclet number of the flow. 
Two types of asymptotic models have been proposed, which yield measurably different predictions for the characteristic vertical velocity and length scale of the turbulent eddies in both diffusive and non-diffusive regimes.  The first, termed a `single-scale model', is designed to describe flow structures having large horizontal and small vertical scales, while the second, termed a `multiscale model', additionally 
incorporates flow features with small horizontal scales, and reduces to the single-scale model in their absence. 
By comparing 
predicted vertical velocity scaling laws with direct numerical simulation data, we show that the multiscale model correctly captures the properties of strongly stratified turbulence within 
\rev{regions dominated by small-scale isotropic motions, whose volume fraction decreases as the stratification increases}. Meanwhile its single-scale reduction accurately describes the more orderly, layer-like, \rev{quiescent} flow outside those 
\rev{regions}.
\end{abstract}



\section{Introduction}

Owing to the associated enhanced rates of irreversible scalar mixing, stratified turbulence is a critical process in the Earth's atmosphere and oceans, impacting both weather and climate, and in the interiors of stars and gaseous planets, affecting their long-term evolution. 
Assuming that the buoyancy of the fluid\rev{, whether liquid or gaseous,} 
is controlled by a single scalar field, which could be temperature or the concentration of a single solute, the dimensionless Boussinesq equations governing the fluid motions \citep{spiegel1960} are
\begingroup
 \allowdisplaybreaks
\begin{subequations}\label{eqn:spiegel_veronis}
\begin{alignat}{2}
     \pd{\boldsymbol{u}}{t} + \boldsymbol{u} \cdot  \boldsymbol{\nabla} \boldsymbol{u}  &= - \nabla p + \frac{b}{Fr^2} \boldsymbol{e}_z + \frac{1}{Re} \nabla^{2} \boldsymbol{u} + \boldsymbol{F}_h, \\
     \pd{b}{t} + \boldsymbol{u} \cdot \boldsymbol{\nabla} b + w &= \frac{1}{Pe} \nabla^2 b, \\
     \boldsymbol{\nabla} \cdot \boldsymbol{u}  &= 0.
 \end{alignat}
 \end{subequations}
 \endgroup
\rev{Here,}~$\boldsymbol{u}=(u,v,w)$ is the velocity field expressed in units of $U\rev{^*}$ (where $U\rev{^*}$ is a characteristic horizontal velocity of the large-scale flow), $t$ is the time variable in units of $L\rev{^*}/U\rev{^*}$ (where $L\rev{^*}$ is a characteristic large horizontal scale of the flow), $p$ is the pressure fluctuation away from hydrostatic equilibrium in units of $\rho\rev{_m^*} U\rev{^{*2}}$ (where $\rho\rev{_m^*}$ is the mean density of the fluid), and $b$ is the deviation of the buoyancy field away from a linearly stratified background, expressed in units of $L\rev{^*} N\rev{^{*2}}$ (where $N\rev{^*}$ is the buoyancy frequency of the stable stratification). \rev{The $w$ term in the buoyancy equation thus represents the vertical advection of the background stratification. Note that here and throughout this paper, starred quantities are dimensional while non-starred quantities are non-dimensional. } The flow is assumed to be driven by a non-dimensional divergence-free horizontal force ${\boldsymbol F}_h$, which only varies on large spatial scales and long time scales. The unit vector $\boldsymbol{e}_z$ points in the direction opposite to gravity. \rev{We note that the validity of the Boussinesq approximation in the context of gaseous atmospheric and astrophysical flows \citep{spiegel1960} must be verified \emph{a posteriori}, by checking that the characteristic vertical scale of the flow remains much smaller than a pressure, temperature or density scale height. It is assumed here that $U^*$ is always much smaller than the sound speed.}

The usual dimensionless governing parameters of the flow emerge; namely, the Reynolds number $Re = U\rev{^*}L\rev{^*}/\nu\rev{^*}$, the P\'eclet number $Pe = U\rev{^*}L\rev{^*}/\kappa\rev{^*}$ and the Froude number $Fr = U\rev{^*}/N\rev{^*}L\rev{^*}$,
where $\nu\rev{^*}$ is the kinematic viscosity, and $\kappa\rev{^*}$ is the buoyancy diffusivity, while the Prandtl number, $Pr =\nu\rev{^*}/\kappa\rev{^*}= Pe/Re$ is a property of the fluid. Typically, $Pr \sim O(1)$  in air and water, but is very small in astrophysical fluids \citep[of order $10^{-2}$ in degenerate plasmas and liquid metals, and much smaller in non-degenerate stellar plasmas, see][]{lignieres2021}. 

 As the stratification increases ($Fr \rightarrow 0$), vertical motions are increasingly suppressed and restricted to small characteristic vertical scales $l_z = O(\alpha)$, where the emergent aspect-ratio $\alpha \rev{ = l_z^* / L^*}$ is an increasing function of $Fr$ but could also depend on $Re$ and $Pe$. In the limit $(\alpha,Fr) \rightarrow 0$, asymptotic analysis has successfully been used to derive reduced equations for stratified turbulence and to gain insight into its properties. \citet{brethouwer2007}, following \citet{billant2001}, proposed an asymptotic reduction in which the vertical coordinate $z$ is rescaled as $\zeta = z / \alpha$ (with $\alpha \ll 1$), while the horizontal coordinates ${\boldsymbol x}_h=(x,y)$ remain of order unity. Accordingly, all dependent variables $q$ are expressed as $q({\boldsymbol x}_h,\zeta,t;\alpha)$. As in \citet{kleinARFM2010}, we refer to this type of model, which is designed to capture the essence of a scale-specific process, as a single-scale asymptotic (SSA) model. 
This vertical rescaling, when used in (\ref{eqn:spiegel_veronis}), reveals the importance of the emergent \emph{buoyancy} Reynolds and  P\'eclet 
 numbers, defined \rev{here} as
 \begin{equation}
 Re_b = \alpha^2 Re \quad \mbox{ and } \quad Pe_b = \alpha^2 Pe,
\end{equation}
 respectively. \citet{brethouwer2007} showed that balancing the mass continuity equation in the limit $\alpha \rightarrow 0$ requires $w = O(\alpha)$. When $Re_b$ and $Pe_b$ are at least $O(1)$, dominant balance in the buoyancy equation implies $b = O(\alpha)$. Finally, the vertical component of the momentum equation reduces to hydrostatic equilibrium when $\alpha \rightarrow 0$, yielding $\alpha = Fr$, as first argued in the inviscid and non-diffusive case by \citet{billant2001}.

More recently, \citet{Shah2024}  noted that in the limit of $Pr \ll 1$, it is possible to have a regime in which $Pe_b \ll 1 \le Re_b$. They demonstrated that in this case, the SSA model and corresponding asymptotic expansion reveal instead that $w = O(\alpha)$ and $b = O( \alpha Pe_b)$, with $\alpha = (Fr^2/Pe)^{1/4}$ 
\citep[see also][]{lignieres2021,Skoutnev2023}.

\begin{figure}
    \centering
    \includegraphics[width=\textwidth]{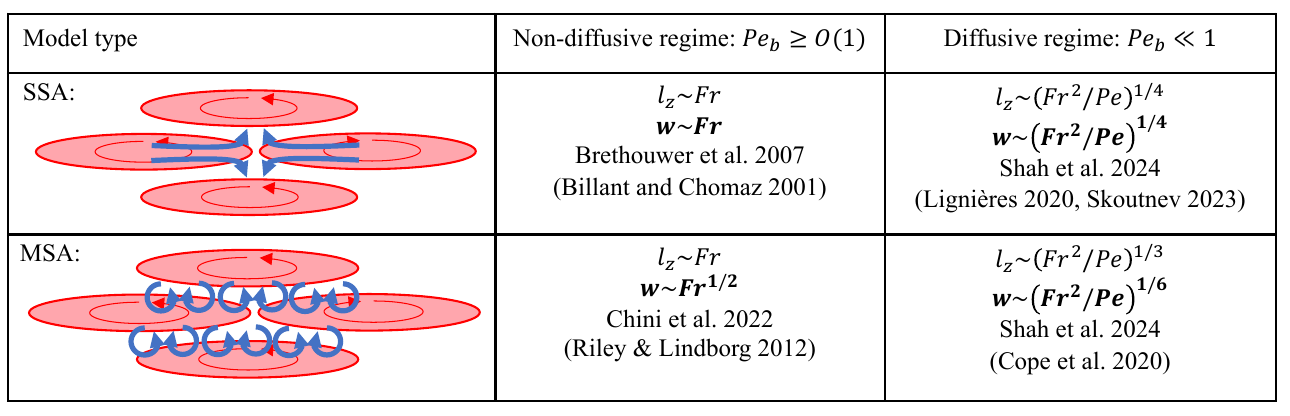}
    
    \caption{Illustrations and summary of the SSA and MSA model predictions for $w$ and $l_z$ in both the non-diffusive and diffusive regimes.  Horizontal eddies are shown in red, and vertical eddies are shown in blue.  }
    \label{fig:comparison-table}
\end{figure}

Crucial to the SSA theory is the notion that {\it every} component of the flow is strongly anisotropic, with large horizontal scales and a small vertical scale. In this model, therefore, the vertical fluid motions are primarily driven by the divergence of the horizontal flow, as illustrated schematically in figure~\ref{fig:comparison-table}. \citet{chini2022}, however, noted that the SSA theory ignores the possibility that \emph{isotropic} motions with small horizontal scales may also exist and, in fact, are commonly seen in numerical simulations of stratified turbulence at sufficiently large $Re_b$ \citep[cf.][]{maffioli2016,cope2020,garaud2020}. They proposed a new asymptotic reduction that explicitly incorporates two horizontal scales and two time scales, such that all dependent variables are expressed as
\begin{equation}
    q({\boldsymbol x}_f,\,{\boldsymbol x}_s,\,\zeta,t_f,t_s;\alpha),\, \mbox{ } \mbox{where } {\boldsymbol x}_s = {\boldsymbol x},\,  {\boldsymbol x}_f = {\boldsymbol x}/\alpha,\, t_s = t,\, t_f = t/\alpha ,
\end{equation}
 where the subscripts $s$ and $f$ are used to denote slow and fast scales, respectively. 
Again following \citet{kleinARFM2010}, we refer to the resulting reduced equations as a multiscale asymptotic (MSA) model. \citet{chini2022} showed that these definitions imply that the large-scale motions remain strongly anisotropic with an aspect ratio $\alpha$, as in the SSA model, but can coexist with isotropic small-scale motions that evolve on the fast time scale $t_f$, and vary on the small vertical coordinate $\zeta$ and the `fast' horizontal coordinate ${\boldsymbol x}_f$.
These small-scale motions are self-consistently driven by an instability of the local vertical shear emergent from the larger-scale horizontal flow (see figure~\ref{fig:comparison-table}), and are gradually stabilized as the stratification increases at fixed $Re$. They essentially disappear beyond a certain threshold, at which point the MSA model naturally recovers the SSA model and its predicted scalings. 
For the sake of clarity, however, we refer in what follows to the SSA model and its scalings whenever small horizontal scales are 
\rev{dynamically negligible},  and to the MSA model and its scalings whenever they are 
\rev{dynamically important}, 
even though the MSA model does in fact naturally cover both cases. 

In the asymptotic limit where $Re_b \ge O(1)$ and $Pe_b \ge O(1)$, 
 \citet{chini2022} found that $w = O(\alpha^{1/2})$, $b = O(\alpha)$ and $\alpha = Fr$ when $\alpha \rightarrow 0$. Their scaling prediction for $w$
 thus deviates substantially from that of \citet{brethouwer2007}, but recovers that of \citet{Riley_Lindborg_2012} albeit using different arguments \citep[for details see][]{Shah2024}. 
 That scaling has been tentatively validated by \citet{maffioli2016} in run-down direct numerical simulations (DNS) of stratified turbulence. 

 Extending the MSA theory to the low $Pr$ case, \citet{Shah2024} recovered the results of \citet{chini2022} when $Pe_b \geq \mathit{O}(1)$. They also found that $\alpha = Fr$ and $w = O(\alpha^{1/2})$ both continue to hold in an `intermediate' regime where $O(\alpha) \le Pe_b \ll 1$. However, when $Pe_b \ll \alpha$, a new fully diffusive regime emerges in which   
 $w = O(\alpha^{1/2})$, $b = O(Pe_b \alpha^{1/2})$ and $\alpha = (Fr^2/Pe)^{1/3}$. As for the $Pr=\mathit{O}(1)$ scenario,
the scaling predictions of the MSA theory differ substantially from those emerging from the low $Pe_b$ limit of the SSA theory 
but recover them when small scales are absent. The various theories and their predicted scalings for $Re_b \ge O(1)$, with $Pe_b \ge O(\alpha)$ or $Pe_b \ll \alpha$, respectively, are summarized in figure~\ref{fig:comparison-table}. 

Therefore, an interesting question is whether evidence for these scaling laws can be found in DNS data. Recently, two series of DNS were presented by \citet{cope2020} and \citet{garaud2020}, respectively, which solved equations (\ref{eqn:spiegel_veronis}) with $\boldsymbol{F}_h = \sin(y) \boldsymbol{e}_x$ (where $\boldsymbol{e}_x$ is a unit vector in the streamwise, i.e.~$x$, direction; see \S\ref{sec:DNS1} for further details). In their $Pe < 1$ simulations (where by construction $Pe_b \ll \alpha$), \citet{cope2020} found that the vertical length scale of the turbulent motions scales as $\alpha = (Fr^2/Pe)^{1/3}$, validating the predictions of \citet{Shah2024} in that limit. This scaling, however, was not as clearly evident in the high $Pe$ but low $Pe_b$ data of \citet{garaud2020}. 
One potential explanation is that $Re_b$ is relatively low in these simulations (which have $Pr = 0.1$, so $Re_b = 10Pe_b$), implying viscous effects are not necessarily negligible. In the limit of high $Pe_b$, \citet{garaud2020} was unable to find evidence for the $\alpha = Fr$, $w = O(Fr)$ scaling of \citet{brethouwer2007} and was unaware at the time of the scaling $w = O(Fr^{1/2})$ obtained by \citet{chini2022}, proposing instead on empirical grounds that $w \propto \alpha = Fr^{2/3}$ provides the best fit to the data. The apparent discrepancy between Garaud's data and previous models therefore prompts us to 
analyze some new DNS results and to
revisit the available data from \citet{cope2020} and \citet{garaud2020} in the light of the MSA models of stratified turbulence recently derived by \citet{chini2022} and \citet{Shah2024}. 
 
\section{Comparison of theory with DNS}
\label{sec:DNS1} 

The SSA and MSA theories differ primarily in their predictions for the characteristic vertical length scale of the flow (or equivalently, $\alpha$) and for the characteristic vertical velocity. We are therefore interested in comparing these predictions to the data. 
In practice, however, the characteristic vertical length scale is a relatively difficult quantity to extract from the DNS, as there is no unique and universally-accepted definition. 
Consequently, we focus solely on comparing the theoretical predictions for the characteristic vertical velocity of the flow to the root-mean-square (rms) of the $w$ field because that quantity is both well-defined and easy to compute. 


In what follows, we extend and re-analyze the datasets presented in \citet{cope2020} and \citet{garaud2020}. 
Both studies performed DNS of the set of non-dimensional equations \eqref{eqn:spiegel_veronis} with $\boldsymbol{F}_h = \sin(y) \boldsymbol{e}_x$ in a triply-periodic domain of size 
\rev{$L_x = 4\pi, L_y = 2\pi, L_z = 2\pi$} using the PADDI code \citep{Traxleral2011}.
\rev{With this choice, $L^*$ is simply the inverse horizontal wavenumber $k^*_f$ of the forcing, and the  dimensional velocity scale $U^*$ is $(F_0^*/\rho^*_m k^*_f)^{1/2}$, where $F^*_0$ is the dimensional amplitude of the forcing. In both studies, the streamwise length of the domain $L_x = 4\pi$ was chosen to be close to the period of the fastest-growing mode of the horizontal shear instability of the Kolmogorov flow driven by that body force \citep[see e.g.][]{cope2020}. This ensures the natural generation of large-scale flows in the horizontal direction, whose nonlinear evolution then generates flows on $O(1)$ scales in both the $x$ and $y$ directions. 
As illustrated in the Appendix, we have verified that 
the results in the non-viscous, non-diffusive regime are essentially independent of $L_x$ and $L_y$ as long as the domain is large enough to allow that primary mode of horizontal shear instability to grow.} 

We focus on the 
highest Reynolds number simulations \rev{of \citet{cope2020} and \citet{garaud2020}}, which were performed for $Re = 600$. Note that \rev{here and in these papers} 
$Re$ is defined using the inverse wavenumber of the forcing, and thus is a factor of $2\pi$ smaller than that of simulations which use the box size as the unit of length instead. \rev{Similarly, $Fr$ is a factor of 2$\pi$ larger here than if we had used the domain size instead.} \citet{cope2020} presented a range of DNS for $Pe \le 0.1$ and low $Fr$ (using the parameter $B = Fr^{-2}$ to characterize the stratification). They also ran a few simulations in the asymptotically low $Pe$ regime \citep{lignieres1999}, called the LPN regime hereafter, where the buoyancy equation is replaced by $w = Pe^{-1} \nabla^2b$. \citet{garaud2020} presented DNS with $Re = 600$, $Pe = 60$ (i.e.~$Pr = 0.1$) and low $Fr$. 
\rev{For each distinct value of $(Re,Pe)$, a first simulation with $Fr = 0.33$ was initialized from $b = 0$ and ${\boldsymbol u} = \sin(y) {\boldsymbol e}_x$, plus small amplitude white noise \citep{cope2020,garaud2020}. Subsequent simulations at higher or lower values of $Fr$ were restarted from the end-point of that first run, to bypass the long transient required for the primary horizontal shear instability to develop.}
Each simulation was integrated until a statistically-stationary state was reached, lasting at least 100 time units \rev{(see the Appendix for sample time-series and a justification for this choice). We have confirmed that the results are independent of the initialization provided the simulations are integrated for at least this period of time.} 
In some cases, we had to further extend the original DNS from \citet{cope2020} or \citet{garaud2020} to have a sufficiently long stationary time series. The quantity $w_{rms} = \langle w^2\rangle_t^{1/2}$, where $\langle \cdot \rangle_t$ denotes a volume and time average, was then measured in that statistically stationary state using the extended data. 

To complement this dataset, we have run additional simulations at $Re = 1000$, $Pe = 100$. These DNS have twice the spatial resolution of those of \citet{cope2020} and \citet{garaud2020} and, thus, have only been integrated for up to 50 time units in the statistically stationary regime. 
In addition, the full fields are too large to be saved regularly, so we have saved two-dimensional slices through the data in the $(x,y)$, $(y,z)$ and $(x,z)$ planes. These simulations are only used for visualizations and to assess the influence of viscosity by comparison with the $Re = 600$ results. \rev{Note that for these new $Re = 1000$ runs, and for all previously published ones in \citet{cope2020} and \citet{garaud2020}, we have ensured that the product of the maximum resolved wavenumber and the Kolmogorov scale is always greater than one, ensuring that the flow field (and the buoyancy field, since $Pr \le 0.1$ in all cases), is resolved down to the dissipation scales \citep[see][for details]{cope2020,garaud2020}.}

\begin{figure}
    \centering
    \includegraphics[width=\textwidth]{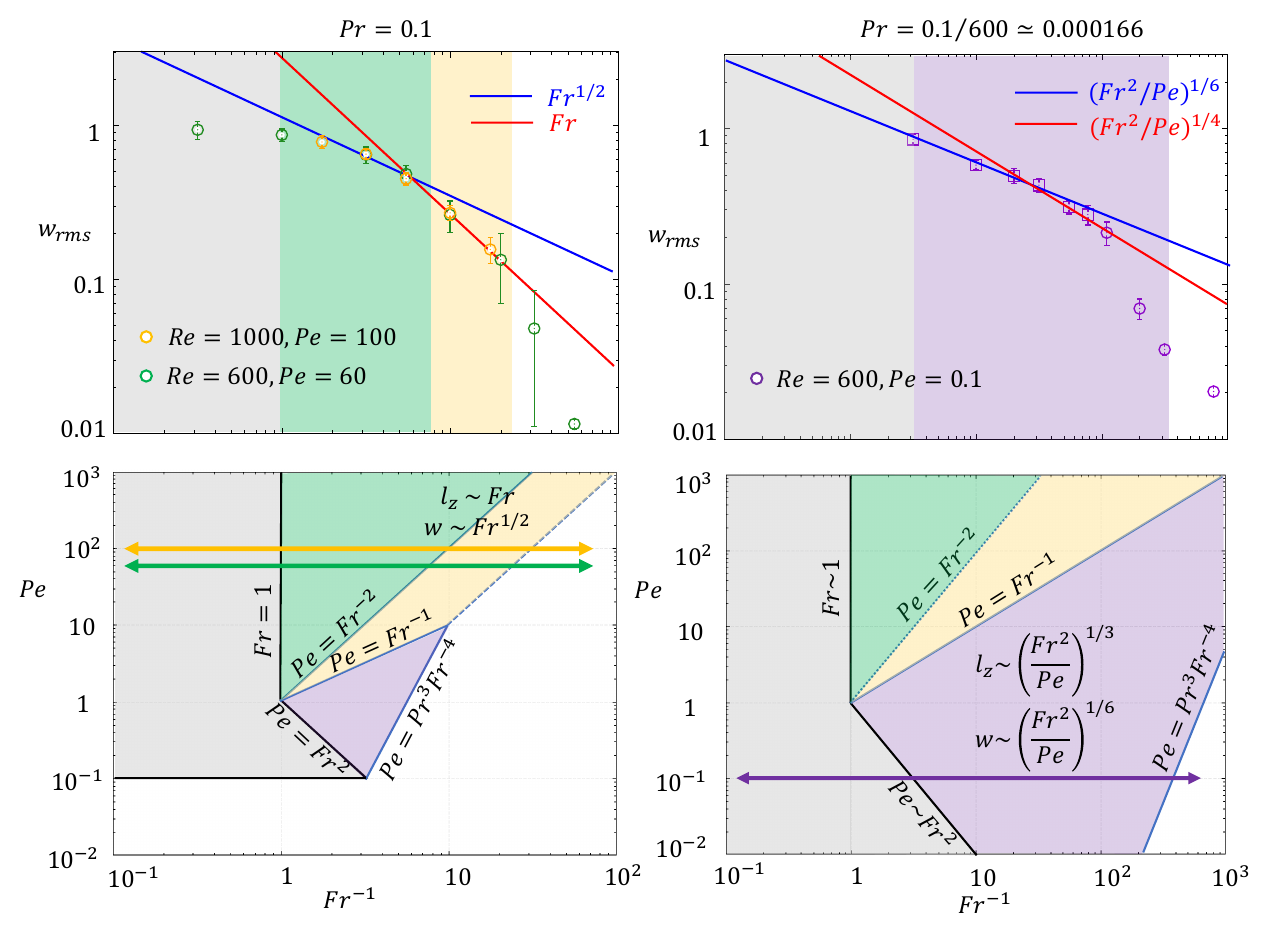}
    \caption{Top row: Comparison between the model predictions and the data for non-diffusive simulations with $Pr = 0.1$ and two different values of $Re$ (left) and  diffusive simulations with $Re = 600$, $Pe = 0.1$ (right). Symbols show $w_{rms}$ extracted from the DNS and errorbars show the standard deviation of its temporal variability. Squares on the right panel denote LPN simulations (see main text for details). The blue and red lines in each panel show 
    the MSA and SSA scaling predictions,
    respectively. Bottom row: Regime diagrams for stratified turbulence at $Pr = 0.1$ (left) and $Pr = 0.1/600 \simeq 0.00017$ (right), adapted from \citet{Shah2024}. Grey regions support isotropic motions. White regions are viscously controlled ($Re_b \le 1$). 
    Green regions support non-diffusive anisotropic stratified turbulence ($Pe_b \ge \mathit{O}(1)$), and purple regions support diffusive anisotropic stratified turbulence ($Pe_b \ll \alpha$).
    The yellow regions are in the `intermediate' regime of \citet{Shah2024} ($\mathit{O}(\alpha) \le Pe_b \ll 1$). Horizontal arrows show the transects through the regimes corresponding to the panels above.}  
    \label{fig:regime_diagram}
\end{figure}

We compare the $w_{rms}$ data and the various theories in the top row of figure \ref{fig:regime_diagram}. The left panel shows $w_{rms}$ as a function of $Fr^{-1}$, measured for the $Pe = 60$, $Re=600$ runs (green symbols), and for the $Pe = 100$, $Re = 1000$ runs (orange symbols). Note that $Pr = 0.1$ in both cases. We refer to these simulations as `non-diffusive' because $Pe$ is large. The fact that the measured values of $w_{rms}$ are identical for the two sets of simulations at different $Re$ demonstrates that viscous effects are negligible, at least for $Fr^{-1} \le 20$. The right panel shows the results of the suite of experiments at $Pe=0.1$, $Re = 600$ (purple symbols), for which $Pr = 0.1/600$. We refer to these simulations as `diffusive' runs, because $Pe$ is small. Note that some of these runs were actually integrated using the LPN regime equations instead (square symbols). In that case, the relevant input parameters are $Re$ and $\chi = Pe/Fr^2$ \citep[$=BPe$ in the notation of][]{cope2020}. To obtain the corresponding value of $Fr$ for a given $\chi$, simply note that $Fr = \sqrt{Pe / \chi}$ for a given $Pe$. Illustrated as well in the same plots are the various theoretical predictions for the vertical velocity: the red line in each panel corresponds to the SSA theory, while the blue line corresponds to the MSA theory.

The bottom row of figure \ref{fig:regime_diagram} shows for comparison the expected regime diagrams for the corresponding values of $Pr$ in each case, based on the asymptotic theory of \citet{Shah2024}. The coloured horizontal arrows show the transect taken through parameter space for each series of DNS shown in the top row. The background colours show the expected regime: isotropic motions with $\alpha \simeq 1$ (grey), non-diffusive anisotropic turbulence (green), intermediate regime (yellow), diffusive anisotropic turbulence (violet) and viscous regime (white). The same background colours in the top row show the expected regime transitions as a function of $Fr^{-1}$ at the value of $Pe$ corresponding to the transect taken. We note that while \citet{Shah2024} distinguished the non-diffusive  and intermediate regimes, these have the same predicted scalings for $\alpha$ and $w$; in any case, the intermediate regime does not span a large region of parameter space at $Pr = 0.1$ and would be difficult to identify even if the scaling laws differed.  

Examination of the top panels confirms that none of the theories applies when the stratification is weak so the flow is isotropic on all scales ($\alpha \simeq 1$, grey region), or in the viscous regime (white region), where $Re_b \le 1$. This is, of course, as expected. However, we also see that neither the SSA nor the MSA model predictions fit the data in the {\it entire} region where they are supposedly valid (i.e. the green/yellow regions in the non-diffusive case, and the purple region in the diffusive case). Instead, we find that the MSA predictions {\it appear} to be better at weaker stratifications (higher $Fr$) while the SSA predictions {\it appear} to be better at higher stratification (lower $Fr$), when $Re$ is fixed. 

\section{
\rev{Turbulent patches vs. quiescent flow}}
\label{sec:DNS2}  

\begin{figure}
    \centering
    \includegraphics[width=\textwidth]{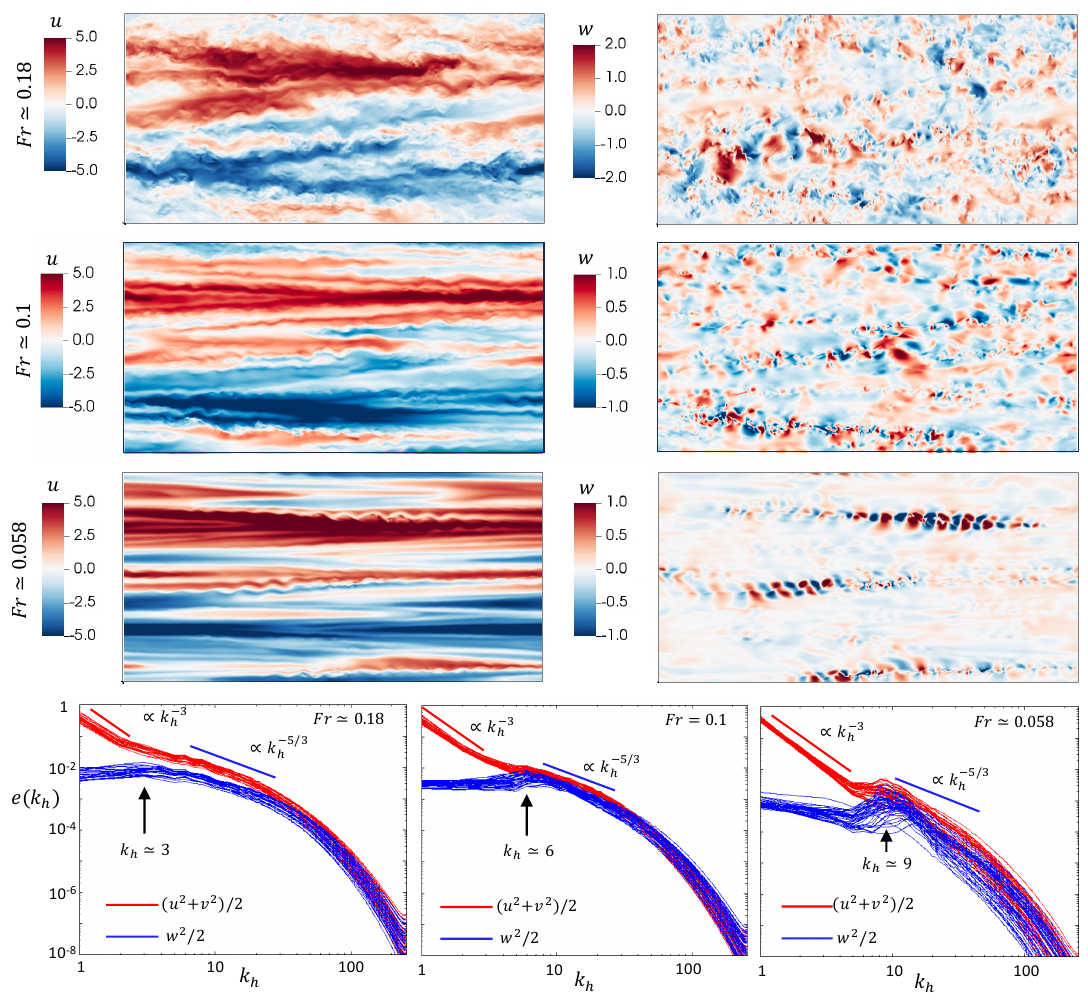}
    \caption{\rev{Top 3 rows:}  DNS snapshots of $u$ 
    and $w$ 
    in the $y=0$ plane for $Re = 1000$, $Pe = 100$ \rev{and various Froude numbers, with stratification increasing from top to bottom}. \rev{Bottom row: Kinetic energy spectra of the horizontal flows (red lines) and of the vertical flows (blue lines) as a function of the horizontal wavenumber $k_h$, for the same three Froude numbers, with stratification increasing from left to right. Each line corresponds to a particular instant in time.} 
    }
    \label{fig:snaps}
\end{figure}

To gain insight into the applicability of the predicted scalings, we examine the actual flow field more closely. \rev{The top three rows of f}igure \ref{fig:snaps} show 
snapshots of $u$ and $w$ in three different high-resolution DNS at $Pe = 100$ and $Re = 1000$\rev{, with $Fr$ decreasing from about $0.18$ to about $0.058$.} 
It is clear that while the $Fr \simeq 0.18$ case is fully turbulent, the more strongly stratified \rev{$Fr \simeq 0.058$} case is not 
as the turbulence is localized to small `patches' (see e.g. the regions of high $|w|$).
Similar findings were reported by \citet{cope2020} in their low $Pe$ simulations in the regime that they named `stratified intermittent' (see their figure 6) \rev{ and are also evident in  \citet{garaud2020}; see the volume renderings of $u$ and $w$ at $Fr = 0.05$ ($B =400$) in her figure 1 for instance. The bottom row of figure \ref{fig:snaps} shows the kinetic energy spectra of the horizontal flows (red lines) and of the vertical flow (blue lines) for the same simulations. Different lines correspond to different instants in time, in order to illustrate the instrinsic variability of the spectra. }

These \rev{results clearly} 
illustrate the coexistence of large and small horizontal scales, with $u$ \rev{increasingly} dominated by large scales with subdominant small scales \rev{as stratification increases}, and $w$ \rev{increasingly} dominated by small scales with subdominant large scales \citep[cf.][]{Riley_Lindborg_2012}. \rev{We see from the spectra in particular that the large scales are highly anisotropic. The horizontal flows have a kinetic energy spectrum proportional to $k_h^{-3}$ (where $k_h$ is the horizontal wavenumber), consistent with oceanic observations \citep[e.g.][]{klymak2007,falder2016} and the classical empirical `Garrett-Munk' spectrum for internal waves \citep{garrett75}. It is also consistent with observations of  turbulence in the stratosphere on $\mathit{O}$(10km) scales \citep{LillyLester1974}.
Meanwhile, the kinetic energy spectrum of vertical motions is a weakly increasing function of $k_h$ at large scales, which peaks at a value of $k_h$ that appears to scale as $Fr^{-1}$ (with the limited data available). This tentatively shows that $l_z^* \simeq Fr L^*$ is indeed the injection scale for the vertical motions in these non-diffusive strongly-stratified simulations, consistent with the interpretation that they arise from  shear instabilities of the layerwise horizontal flow on that scale. 

At smaller scales, we see that the flow becomes much more isotropic. In the $Fr \simeq 0.18$ and $Fr =0.1$ cases, for which $Re_b = Fr^2 Re \simeq 33$ and $10$, respectively, the isotropic small-scale flow seems to have a standard $k_h^{-5/3}$ spectrum  until viscous effects come into play. In the $Fr \simeq 0.058$, by contrast, the spectrum of the small-scale flow is steeper than $k_h^{-5/3}$, indicating that the turbulence is somewhat suppressed. 
This observation is not surprising, because at these parameter values $Re_b \simeq 3.3$, and so viscous effects are likely to suppress inertial range dynamics that are unaffected by stratification, as well as suppressing the local vertical shear instability except in regions where the shear is exceptionally strong, leading to the patchiness of the flow observed in the snapshots.
}

\rev{The snapshots} in the more strongly stratified case \rev{further reveal} that the 
small horizontal scales are only dominant within the turbulent patches and essentially disappear outside of these patches. As such, the distinct MSA model scalings are only expected to apply within the turbulent patches. In the more orderly layer-like flow outside of those patches, the SSA scalings---which coincide with the MSA model predictions in regions where small scales are not excited---should hold.

\begin{figure}
    \centering
    \includegraphics[width=0.9\textwidth]{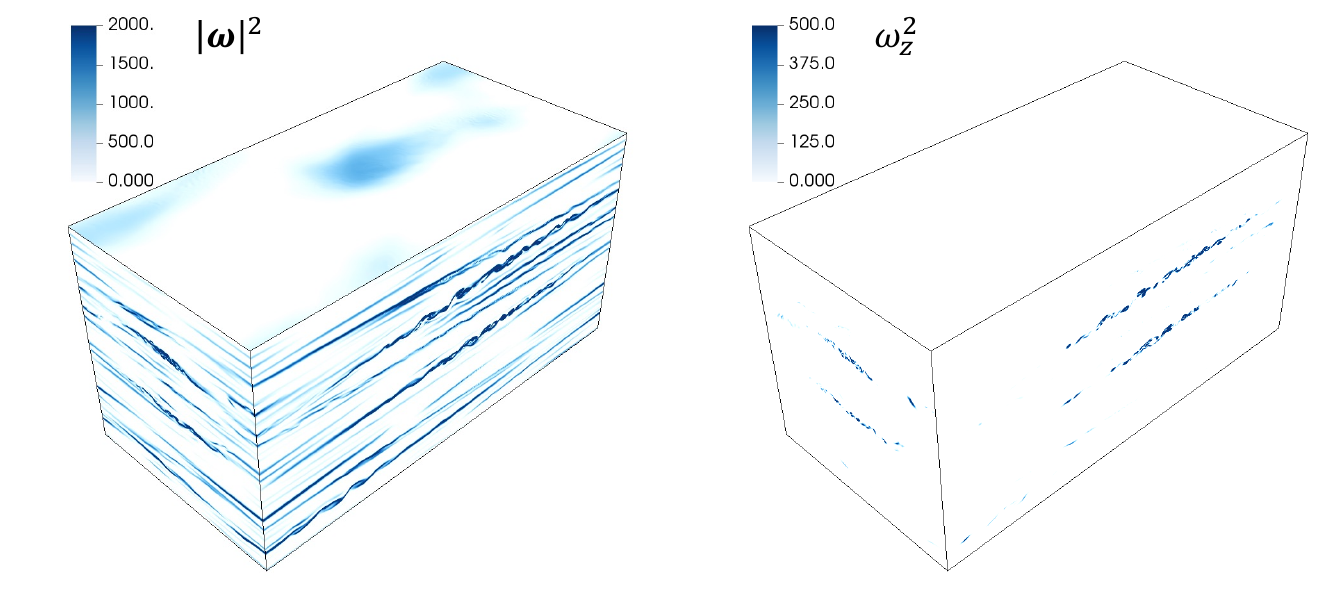}
    
    \caption{Snapshots of enstrophy (left) and vertical vorticity squared (right) from a simulation at $Re = 600$, $Pe = 60$, and $Fr = 0.05$. The latter is a better diagnostic of the turbulent patches.
    }
    \label{fig:enstrophy}
\end{figure}

To verify this interpretation quantitatively, we sought to identify a reliable diagnostic for the turbulent patches, i.e. regions where the flow exhibits  small horizontal scales. It is common to use the enstrophy $|{\boldsymbol{\omega}}|^2$ as a diagnostic for turbulence, where $\boldsymbol{\omega} = \nabla \times {\bf u}$ is the flow vorticity. Indeed, the turbulent cascade to small scales implies that enstrophy must be large within the patches. However, enstrophy turns out to be an inappropriate diagnostic for our purpose because it can {\it also} be large in the layer-like regions of strong vertical shear outside of the turbulent patches, such as the ones described by the SSA model. 
This fact is illustrated in figure \ref{fig:enstrophy}(a), which shows the enstrophy field in a particular snapshot of a strongly stratified simulation, and can be understood as follows. According to \citet{chini2022} and \citet{Shah2024}, ${\boldsymbol u} = \bar {\boldsymbol u} + {\boldsymbol u}'$ where $\bar {\boldsymbol u}$ can be thought of as the large-scale anisotropic component of the flow, which varies on the $O(1)$ horizontal scales and $O(\alpha)$ vertical scale, as in the SSA model. Meanwhile ${\boldsymbol u}'$ can be thought of as the small-scale isotropic and turbulent component of the flow, which varies on $O(\alpha)$ scales in all directions, as in the MSA model. Furthermore, these authors show that $\bar u \sim \bar v \sim O(1)$, while $\bar w \sim O(\alpha)$, and $u'\sim v' \sim w' \sim O(\alpha^{1/2})$. Accordingly, we find that the horizontal vorticity components are dominated by the contribution from $\bar {\boldsymbol u}$, namely $\omega_x \sim \bar \omega_x \sim \omega_y \sim \bar \omega_y \sim O(Fr^{-1})$, while the vertical vorticity component is dominated by the contributions from ${\boldsymbol u}'$, with $\omega_z \sim \omega'_z \sim O(Fr^{-1/2})$. \rev{For comparison, note that the vertical vorticity associated with the large-scale forcing, and with the mean horizontal flow $(\bar u, \bar v)$, is $O(1)$ and therefore negligible in comparison with $\omega_z'$.} As a result, we argue that $\omega_z^2$ is a more reliable diagnostic of the small-scale turbulence than the enstrophy. 
This assertion is confirmed in figure~\ref{fig:enstrophy}(b), which shows $\omega_z^2$ for the same snapshot depicted in figure~\ref{fig:enstrophy}(a). We see that the regions of high $\omega_z^2$ {\it only} highlight the turbulent patches of the flow.  

In what follows, we therefore define the following quantities\rev{, using weighted averages with the weight function $\omega_z^2$
to emphasize the turbulent patches and the function $\omega_z^{-2}$ to emphasize the more quiescent regions, respectively}: 
\begin{equation}
    w^{\rm turb}_{rms} = \frac{\langle w^2 \omega_z^2\rangle_t^{1/2}}{\langle \omega_z^2\rangle_t^{1/2}},  \quad \mbox{ and }\quad  w^{\rm no turb}_{rms} = \frac{\langle w^2 \omega_z^{-2}\rangle_t^{1/2}}{\langle \omega_z^{-2}\rangle_t^{1/2}  }.
\end{equation}
The first can 
be viewed as the \rev{rms} of $w$ taken over the turbulent patches, where the distinct MSA scalings should apply. The second can be viewed as the rms of $w$ taken everywhere {\it other} than the turbulent patches, where the SSA scalings should apply. Note that the computation of $w^{\rm turb}_{rms}$ and $w^{\rm noturb}_{rms}$ requires integrals of $w^2$,  $\omega_z^2$, and their product or ratio over the entire volume, which was not one of the simulation diagnostics originally saved. As such, we are unable to extract these quantities from the $Re = 1000$ simulations. However, we can compute them from the full-data snapshots regularly saved in the  $Re = 600$ simulations in both high and low $Pe$ datasets (of which there are usually between 50 and 100 depending on the simulation). The variance is naturally larger than for the $w_{rms}$ data, \rev{because of the smaller amount of data available.} 
\rev{We also note that similar results can be obtained by choosing $|\omega_z|$ and $|\omega_z|^{-1}$ as the weight functions for the turbulent and quiescent regions, respectively. However, $|\omega_z|$ is only a factor of $Fr^{-1/2}$ larger in the turbulent regions than in the quiescent ones. Since $Fr$ in our simulations is not extremely small, we use $\omega_z^2 = O(Fr^{-1})$ instead to more clearly identify the turbulent patches.}

\begin{figure}
    \centering
\includegraphics[width=\textwidth]{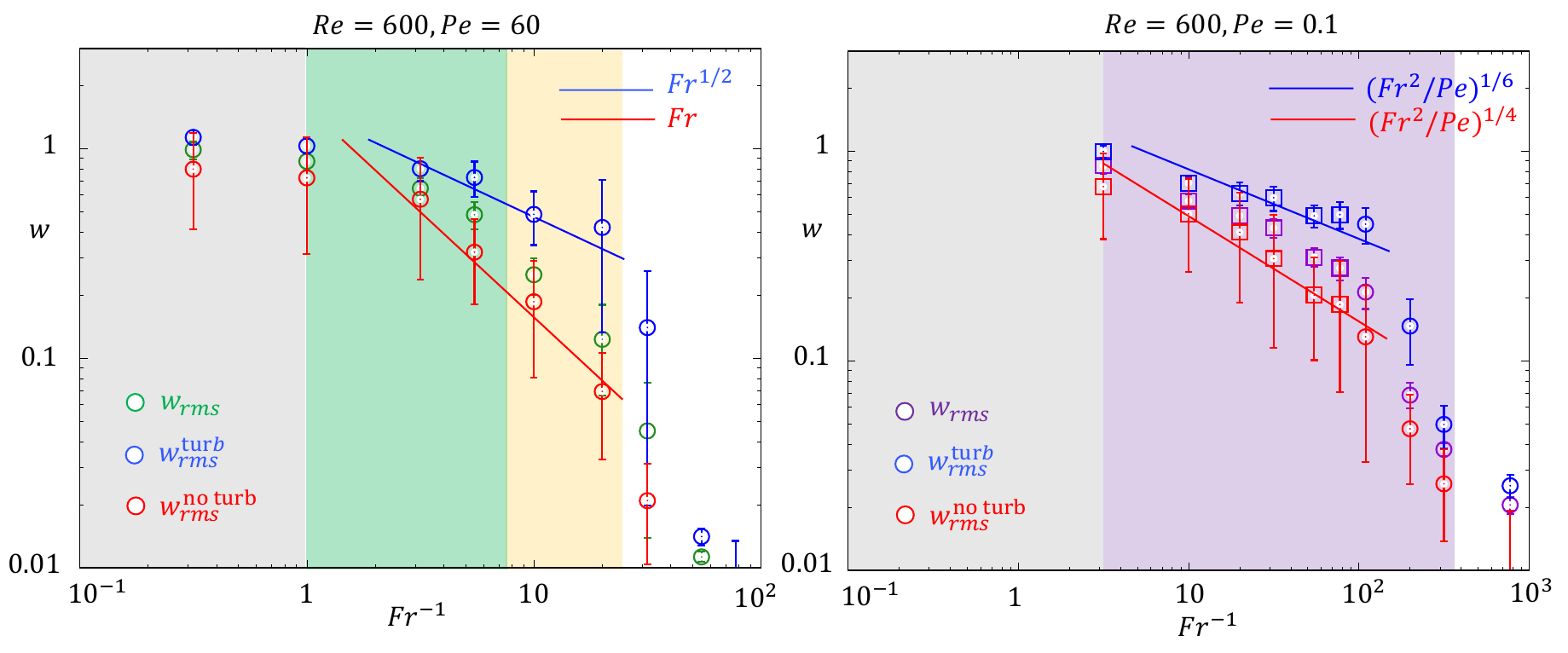}
    \caption{Comparison between models and data for the characteristic vertical velocity at $Re = 600$, $Pe = 60$ (left) and at $Re = 600$, $Pe = 0.1$ (right). Green and purple symbols show $w_{rms}$ in the left and right panels, respectively. In both panels, blue symbols show $w^{\rm turb}_{rms}$ and should be compared with the turbulent MSA scalings (blue lines), while red symbols show $w^{\rm noturb}_{rms}$ and should be compared with the corresponding SSA scalings (red lines). 
    }
 \label{fig:tadaa}
\end{figure}

We present the results in figure~\ref{fig:tadaa}, with the non-diffusive $Re = 600, Pe =60$ simulations on the left and the diffusive $Re = 600, Pe = 0.1$ simulations on the right. The background colours are the same as in figure~\ref{fig:regime_diagram}. The $w_{rms}$ data from figure~\ref{fig:regime_diagram} is again shown in green and purple symbols. We plot the $w^{\rm turb}_{rms}$ data using blue symbols in both cases, and the MSA scalings for turbulent regions using a blue line for comparison. Similarly, we plot the $w^{\rm noturb}_{rms}$ data using red symbols and the SSA scalings using a red line. We see, quite clearly, that each theory fits the data in its respective region of validity---the distinct MSA scalings being valid in the turbulent patches, and the SSA scalings only being valid outside of the turbulent patches. 
This shows that the transition observed in figure~\ref{fig:regime_diagram}, from simulations that appear to satisfy the turbulent MSA scalings better at low stratification to simulations that appear to fit the SSA scalings better at high stratification, primarily is a consequence of the decrease in the volumetric fraction of the domain occupied by the turbulent patches when $Fr^{-1}$ increases. 

As the stratification continues to increase, the buoyancy Reynolds number $Re_b = \alpha^2 Re$ eventually decreases below a critical value $Re_{b,{\rm crit}} = O(1)$, where viscous effects become dominant. Assuming  $Re_{b,{\rm crit}} =1$, we show this transition in figure \ref{fig:regime_diagram} as the line separating the coloured area from the white region, for the MSA model. In the non-diffusive case (left panel), $\alpha = Fr$, so $Re_{b} =1$ is equivalent to $Pe = Pr Fr^{-2}$, which is the edge of the yellow region. We see that the data are consistent with this prediction: beyond the viscous transition, $w_{rms}$ rapidly drops to very low values consistent with a viscously-dominated flow. For the diffusive case (right panel), $\alpha = (Fr^2/Pe)^{1/3}$ in the MSA model so $Re_b = 1$ is equivalent to $Pe = Pr^3 Fr^{-4}$, which corresponds to the edge of the purple region. We see that 
the effects of viscosity appear to become important at slightly weaker stratification than predicted assuming $Re_{b,{\rm crit}} =1$, but plausibly attribute this discrepancy to missing $O(1)$ constants in the estimates for $\alpha$ and/or $Re_{b,{\rm crit}}$.

\section{Conclusion}
In this paper, we have presented a detailed comparison of DNS data with various theoretical predictions for the characteristic vertical velocity of fluid motions in forced stratified turbulence. In particular, we have studied both moderate and low Prandtl number regimes, resulting in a wide range of P\'eclet numbers at fixed Reynolds number. When buoyancy diffusion is negligible, our results notably provide compelling evidence for the $w \propto Fr^{1/2}$ scaling law for stratified turbulence first proposed by \citet{Riley_Lindborg_2012} using heuristic arguments and rigorously derived by \citet{chini2022} using multiscale asymptotic analysis. In the latter investigation, this scaling law is intrinsically tied to the existence of small-scale isotropic flow motions driven by the emergent vertical shear between larger-scale primarily horizontal eddies, as corroborated by the results from the DNS presented here. The vertical shear instability is gradually stabilized as the buoyancy Reynolds number decreases towards a critical value of order unity, and the small-scale isotropic component of the turbulence becomes 
\rev{confined to localized patches} rather than \rev{being} domain filling. Outside of the\rev{se} turbulent patches, small horizontal scales disappear, and we find that $w \propto Fr$ instead, consistent with the model of  \citet{billant2001} and \citet{brethouwer2007}. 
\rev{As $Re_b$ gradually approaches unity from above}, therefore, the rms vertical velocity of the flow computed from an average over the {\it whole} domain differs from either of these scaling laws, and additionally depends on the volume filling factor of the small-scale turbulence, whose dependence on stratification and $Re_b$ \rev{is the subject of ongoing work}. 
\rev{The patchiness of the small-scale isotropic flow} also explains the incorrect conclusion reached by \citet{garaud2020} regarding the possible existence of another regime of stratified turbulence where $w \propto Fr^{2/3}$. In hindsight, we understand her empirically-inferred \rev{intermediate} scaling as a consequence of the decrease in the volume filled by turbulent patches with increasing stratification at fixed $Re$.  

At low $Pr$, \citet{Shah2024} revised the predictions of \citet{brethouwer2007} and \citet{chini2022} to account for the effects of buoyancy diffusion. They showed that the presence of small isotropic motions implies that $w \propto (Fr^2/Pe)^{1/6}$, consistent with an early model and DNS data by \citet{cope2020}, while in their absence $w \propto (Fr^2/Pe)^{1/4}$, consistent with predictions from \citet{lignieres2021} and \citet{Skoutnev2023}. Revisiting the very low $Pr$ DNS of \citet{cope2020} in this new light, we have confirmed both scaling laws within and outside of the turbulent patches, respectively.
Finally, the models of \citet{chini2022} and \citet{Shah2024} also predict where in parameter space viscous effects become important. We have confirmed these predictions, too, with our DNS data. 

\rev{An important caveat of our conclusions is that the simulations were limited to a specific type of horizontal forcing. We believe that the results ought to apply more generally as long as the forcing drives primarily horizontal flows on long time scales and large length scales, but this conjecture will need to be verified in future work.}

In summary, this investigation demonstrates that the combination of rigorous multiscale analysis \citep{chini2022,Shah2024} with idealized DNS \citep[][and new simulations presented here]{cope2020,garaud2020} can be a powerful tool to identify and validate scaling laws for stratified turbulence across different regions of parameter space. In future work, we will incorporate the effects of rotation and magnetic fields, which must be taken into account for a more realistic description of stratified turbulence in geophysical and astrophysical settings.

\section*{Acknowledgements}

\noindent \rev{This work uses the Expanse supercomputer at the San Diego Supercomputing Center. P. Garaud thanks the SDSC support team for their help}. The authors gratefully acknowledge the Geophysical Fluid Dynamics Summer School (NSF 1829864), particularly the \rev{2018}, 2022 and 2023 programs.
K.S. acknowledges funding from the James S. McDonnell Foundation. G.P.C. acknowledges funding from the U.S. Department of Energy through award DE-SC0024572.
For the purpose of open access, the authors have applied a Creative Commons Attribution (CC BY) licence to any Author Accepted Manuscript version arising from this submission.

\section*{Declaration of Interests}

\noindent The authors report no conflict of interest.

\section*{\rev{Appendix: Numerical considerations}}

\begin{figure}
    \centering
    \includegraphics[width=\textwidth]{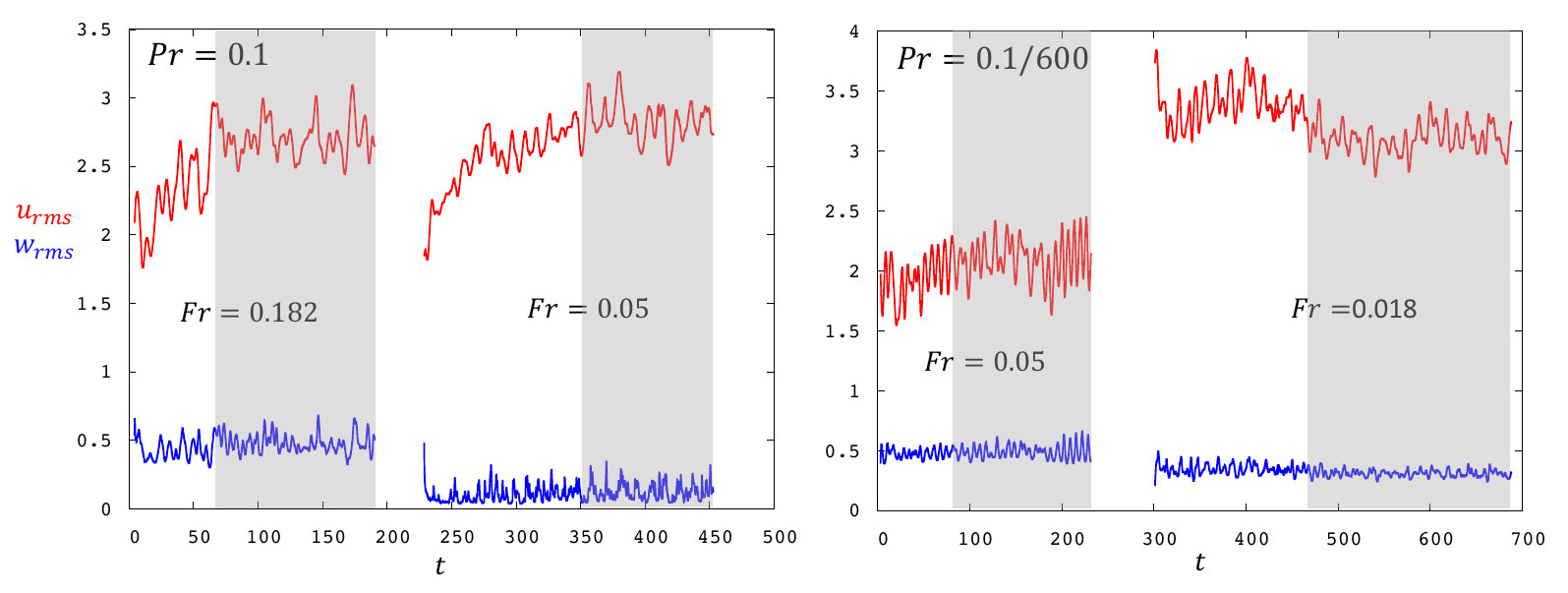}
    
    \caption{\rev{Sample time series of the instantaneous rms horizontal velocity $u_{rms}$ (red lines) and vertical velocity $w_{rms}$ (blue lines), as a function of time for various simulations.  Note that the starting points of the simulations have been offset to an arbitrary position for ease of visualization. The grey area shows the time-averaging interval used in figures \ref{fig:regime_diagram} and \ref{fig:tadaa}}.
    }
    \label{fig:app1}
\end{figure}

\rev{The results presented in this paper require simulations that have achieved a statistically stationary state lasting at least 100 time units. This duration was chosen to ensure that the time series from which flow statistics are computed are sufficiently uncorrelated in time. Indeed, assuming a mean streamwise velocity of 1, the fluid has time to flow approximately 8 times through the domain in 100 time units when the domain length is $4\pi$. In practice, the mean streamwise velocity ranges from about 2 to 4, depending on the input parameters, and the integration time interval is often larger than 100 time units, so the true number of `laps' is generally much higher. Time series of the instantaneous rms streamwise and vertical velocity for a few selected  simulations in both non-diffusive and diffusive regimes are shown in figure \ref{fig:app1}. }

\begin{figure}
    \centering
    \includegraphics[width=0.9\textwidth]{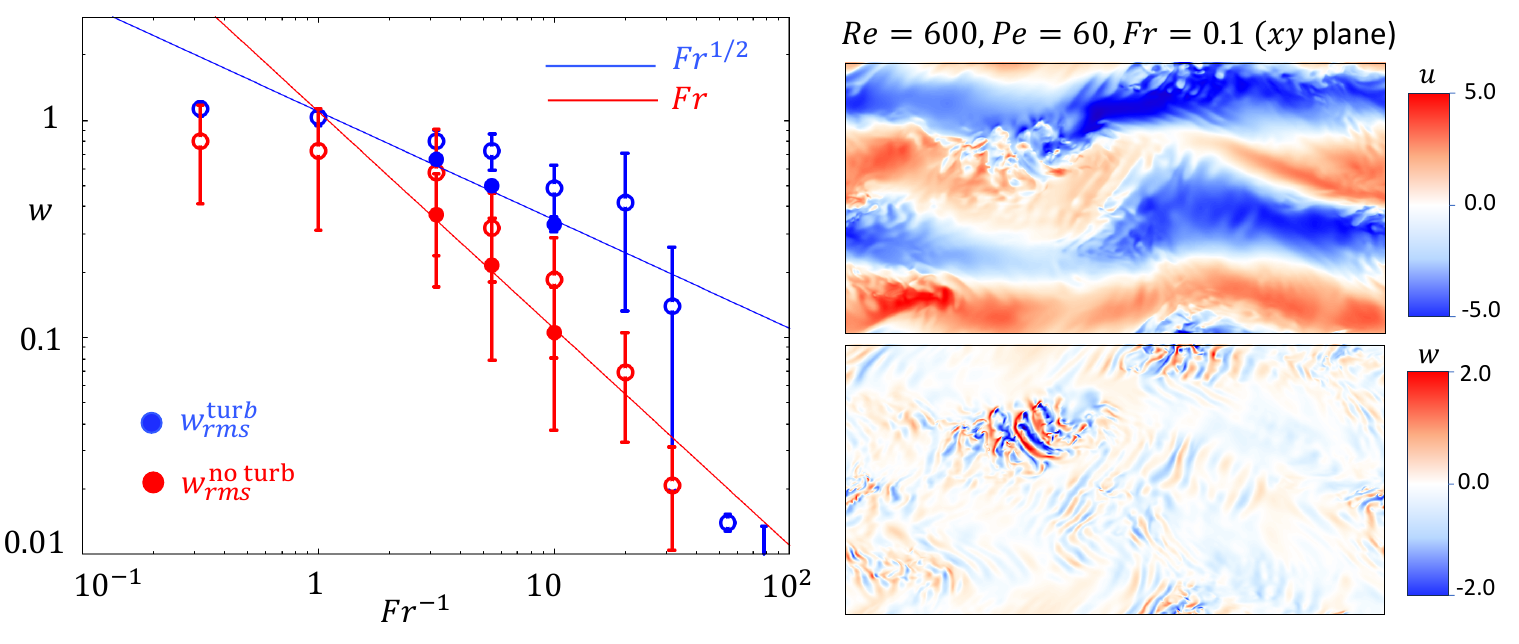}
    
    \caption{\rev{Left: Filled symbols show $w^{\rm turb}_{rms}$ and $w^{\rm noturb}_{rms}$
    extracted from simulations in larger computational domains (of size $8\pi \times 4\pi \times 2\pi$) and can be compared to those extracted from simulations in regular-sized domains, shown as open symbols. Right: Snapshot of the $(x,y)$ plane at some arbitrary value of $z$ in the statistically stationary state of a simulation at $Re = 600$, $Pe = 60$, $Fr = 0.1$, in a domain of size $8\pi \times 4\pi \times 2\pi$.}  }
    \label{fig:app2}
\end{figure}

\rev{We have also verified that the scaling laws obtained are independent of the selected domain size by performing a few simulations in a $8\pi \times 4\pi \times 2\pi$ domain with the same body force. With that choice, the domain is sufficiently wide to accommodate two wavelengths of the applied sinusoidal force in the spanwise direction, and two wavelengths of the fastest-growing mode of horizontal shear instability in the streamwise direction. Because of the heavily increased computational cost, we have only run cases for $Re = 600, Pe = 60$, and three values of $Fr$, and these simulations have been run for a shorter duration. Yet, as demonstrated in figure \ref{fig:app2}, the same scaling laws are found in the turbulent patches ($w^{\rm turb}_{rms} \propto Fr^{1/2}$) and outside of the turbulent patches ($w^{\rm no turb}_{rms} \propto Fr$), respectively. We note that the prefactor is slightly smaller, but this is not surprising given that the meandering flow structure is allowed to be a little different in the larger domain. 
}

\bibliographystyle{jfm}

\end{document}